# Precision Profile Weighted Deming Regression for Methods Comparison

Douglas M Hawkins,[1]

Jessica J Kraker[2]

1. School of Statistics, University of Minnesota, Minneapolis, MN, USA.
2. Mathematics Department, University of Wisconsin, Eau Claire, WI, USA.

***Running title*:** Precision profile weighted Deming methods comparison

**Abstract**

*Background*

Validating a test method is commonly done by methods comparison (MC) using an accepted predicate to measure samples and to assess their agreement.  The most powerful way of doing so is by parametric modeling: an "errors-in-variables", or Deming, regression.  The measurement variance is rarely constant across the measuring interval, and so Deming regression requires suitable weights to achieve its theoretical good performance.  Sometimes, independently of the data set, we have an explicit mathematical model, the precision profile, connecting each assay's variability to the analyte concentration.  But many studies have no such external information and the analysis must rely on the data set alone, with some assumption about the form of the precision profile.

*Results*

Weighted Deming regression is outlined, and R codes for implementing it are provided in the known and unknown precision profile settings. The implementation includes a jackknife approach for standard errors and confidence intervals for the regression parameters, residuals for diagnostics on normality and linearity, and a methodology for identifying and testing outliers.

*Conclusions*

Existing weighted Deming regression publications assume either constant coefficient of variation or constant variance. Precision profile models fill this gap and allow for more general settings.

**Impact statement**

When the underlying Gaussian model is appropriate, weighed Deming regression is the statistically efficient preferred method of analysis. It also has the benefit of providing residuals with known statistical distribution. The R codes provided make it generally accessible to a broad class of potential users.

*Introduction*

Methods comparison is a staple problem in clinical chemistry. In a common framework, a new "test' assay is to be compared with a "predicate" assay, but the statistically identical problem comes up in, for example, comparison of tube types, some approaches to storage stability, and lot calibration.

The relationship between the two assays is captured in a regression-like equation, the most common special case of which is a linear regression. As both *X* and *Y* incur measurement random variability, conventional least squares regression is inappropriate. "Errors in variables" techniques accommodate this and have a long history of use in econometrics. There, however, it is assumed that the measurement variability is constant across the range of data. This is rarely true in chemical assays spanning an order of magnitude or more. Here two regression methods predominate [1] – the nonparametric Passing Bablok; and Linnet's constant coefficient of variation Deming.

Passing Bablok has no distributional assumptions and is outlier-resistant. It is generally agreed to work well in very highly correlated data, but less so with lower correlation and where the assays have different precisions. Also its lack of distributional assumptions translates into skimpy tools for analyzing residuals, which is an important part of any regression analysis. Turning to Deming, many assays have constant coefficient of variation and reasonably normal measurement errors. Linnet's implementation of this setting however suffers from an implicit assumption that the true regression is close to the line of identity. As shown in the supplemental appendix, when this is not true its estimates are biased.

This leaves an opening for statistically efficient parametric methods that are not tied to either the constant variance or the constant CV assumptions; that implement the model correctly whatever the true regression line; and that provide good residual diagnostics. It is also important to have software to perform the non-trivial calculations involved. That is the scope of this proposal.

*General weighted Deming regression*

Suppose we have *n* samples with true values $\mu_i$  $i = 1, 2, \ldots, n$ measured by a test method *Y* and an accepted predicate method *X*. The predicate assay is assumed to be unbiased for the true value and to follow a Gaussian distribution with a variance dependent on the mean.

$$X_i \sim N[\mu_i, g(\mu_i)] \tag{1}$$

The precision profile is an explicit mathematical function for *g*. The simplest choices are

- $g(\mu_i) = \sigma^2$, the constant variance model,
- $g(\mu_i) = (\kappa \mu_i)^2$ the constant coefficient of variation model

Combining these gives the Rocke-Lorenzato (RL) model

$$g(\mu_i) = \sigma^2 + (\kappa \mu_i)^2 \tag{2}$$

These, and a variety of other precision profile models are described in the standalone VFP program [2] and its R counterpart [3], both of which may be used to fit these models to actual data. We return to this point later.

Similarly, the model for the test method is

$$Y_i \sim N[\alpha + \beta \mu_i, h(\alpha + \beta \mu_i)]$$

To simplify notation, abbreviate the variances to $g_i$ and $h_i$ ignoring their dependence on $\mu, \alpha, \beta$. Omitting the irrelevant constant, -2 log likelihood is given by

$$L = \sum_{i=1}^{n} \left[ \frac{(X_i - \mu_i)^2}{g_i} + \frac{(Y_i - \alpha - \beta\mu_i)^2}{h_i} + \log(g_i) + \log(h_i) \right] \qquad (3)$$

The parameters are the regression slope and intercept $\alpha, \beta$ and the underlying unknown latent true values $\mu_i$. Differentiating with respect to $\mu_i$ gives:

$$\frac{\partial L}{\partial \mu_i} = \frac{-2(X_i - \mu_i)}{g_i} - \frac{2\beta(Y_i - \alpha - \beta\mu_i)}{h_i} \qquad i = 1, 2, \ldots n$$

Setting this to zero gives the maximum likelihood estimators

$$\mu_i = \frac{h_i X_i + g_i \beta(Y_i - \alpha)}{h_i + g_i \beta^2} \qquad (4)$$

The slope and intercept are conceptually given by the least squares regression of $Y$ on these intermediate latent quantities $\mu$ with weights $1/h_i$ but are generally found using formulas that hide this interpretation.

The R function PWD_known implements this calculation. The user supplies $X, Y,$ and functions to calculate $g$ and $h$, and the code returns the estimated $\mu, \alpha, \beta$ and the minimized -2 log likelihood.

Following standard statistical practice, we will use "hat" notation for estimates, ie $\hat{\alpha}, \hat{\beta}, \hat{\mu}_i$. Residuals are a central feature of model checking -- an intuitive residual is

$$e_i = Y_i - \hat{\alpha} - \hat{\beta} X_i \qquad (5)$$

Ignoring the second-order contribution from the random variability in $\hat{\beta}$ (justified by the large sample typically used in MC studies) and plugging in its estimate, the variance of the residuals is

$$\text{Var}(e_i) = h_i + g_i \beta^2, \qquad (6)$$

Having fitted the regression model one can then compute the residuals and rescale to a putatively standard normal scaled residual

$$r_i = e_i / \sqrt{h_i + g_i \hat{\beta}^2} \tag{7}$$

These should be plotted to see if there is evidence of curvature, implying a nonlinear relationship between the instruments; of nonconstant variance, implying a wrong specification for *g* and/or *h*; and of apparent outliers. They can also be tested for normality.

Returning to *g* and *h*, there are two potential sources of estimates. If the assays *X* and *Y* are in replicate, then modeling the variance of each set of replicates against their mean can provide estimates of *g* and *h*. The standalone or the R versions of VFP [2,3] provide suitable tools for this modeling. As the MC then uses the means of the replicates, the variances of *g* and *h* must be divided by the number of replicates if the residuals are to be scaled correctly.

Alternatively, if *X* and/or *Y* is singlicate but a separate precision study of the assay is available, then that precision study could be analyzed to get the required *g* and *h*.

Both sources should be treated with some caution as they may not reflect all relevant sources of variability in the actual assay pairs. For this reason, it is advisable to check that the scaled residuals have standard deviation of 1, as they should.

*Standalone weighted Deming*

Frequently the assays are singlicate and there is no plausible external source for *g* and *h*. In this setting, options are limited and there is no prospect of estimating *g* and *h* separately. If it is plausible that *h=g*, then a consensus common variance profile model can be fitted using singlicate data as described in [4]. That paper assumed the RL model

$$g = \sigma^2 + [\kappa\mu]^2$$

$$h = \sigma^2 + [\kappa(\alpha + \beta\mu)]^2$$

a modest extension of which is

$$g = \lambda(\sigma^2 + [\kappa\mu]^2)$$

where the constant of proportionality $\lambda$ is estimated externally in some way.

Under this model, ignoring the contribution of the externally-specified constant $\lambda$ to the log variance term, the -2 log likelihood is

$$L = \sum_{i=1}^{n}\left[\frac{(X_i - \mu_i)^2}{\lambda(\sigma^2 + \kappa^2\mu_i^2)} + \frac{(Y_i - \alpha - \beta\mu_i)^2}{\sigma^2 + \kappa^2(\alpha + \beta\mu_i)^2} + \log\{(\sigma^2 + \kappa^2\mu_i^2)(\sigma^2 + \kappa^2(\alpha + \beta\mu_i)^2)\}\right] \quad (8)$$

The maximum likelihood (ML) fit is obtained by minimizing this function over the two variance profile parameters $\sigma, \kappa$, the two regression parameters $\alpha, \beta$, and the $n$ nuisance parameters $\mu_i$.

The maximum likelihood estimators for the profile parameters turn out to have a substantial bias. This is illustrated by a simulation of 1,000 replicates of a data set with $n=100$ samples having true $\mu$ values in geometric progression from a low of 20 to a high of 100 units. The $X$ and $Y$ values were generated using $\alpha=0$, $\beta=1$ and the RL precision profile model with $\sigma=5$, $\kappa=0.1$. The top half of Figure 1 shows box and whisker plots of the variance profile parameter estimators with the true values as horizontal lines. Clearly the estimates of both $\sigma$ and $\kappa$ have major bias.

This bias does not in and of itself bias the regression parameters; these merely require that the estimated variance profile be a constant multiple of the true profile. The lower half of Figure 1 compares the intercept and the slope fitted using the true $\sigma, \kappa$ with that using the estimated $\sigma, \kappa$.

The two are close to identical. Summary statistics and Z scores testing for bias are:

|      | Intercept |           | slope  |           |
|------|-----------|-----------|--------|-----------|
|      | known     | estimated | known  | estimated |
| mean | -0.098    | -0.090    | 1.002  | 1.004     |
| sd   | 2.332     | 2.359     | 0.051  | 0.051     |
| Z    | -4.202    | -3.815    | 3.922  | 7.843     |

All four estimators have statistically significant but practically inconsequential biases. Crucially, the regression estimates based on the fitted variance profile model have virtually the same performance as the utopian known-parameter regression despite the large biases in the fitted variance profiles. The conclusion is that having to estimate the precision profile rather than use its exact value has no perceptible cost in either bias or efficiency. This general conclusion has been confirmed in simulations of a variety of $\sigma, \kappa$ values.

*Inference*

The standard errors of the estimators $\hat{\alpha}, \hat{\beta}$ and their covariance are conventionally estimated by the jackknife or the bootstrap [5]. The R function PWD_inference implements this consensus approach and provides the estimates of the intercept and slope along with their standard errors and covariance. It uses the RL variance profile model.

*Residuals*

Under the RL model, the residuals have variance $\lambda\beta^2(\sigma^2 + [\kappa X]^2) + \sigma^2 + [\kappa(\alpha + \beta X)]^2$ which in principle could be estimated by plugging in the estimates of $\sigma, \kappa, \alpha, \beta$ and the externally-supplied value of $\lambda$ and ignoring the uncertainty in the parameter estimates. But as the estimates of $\sigma, \kappa$ from optimizing (8) are seriously biased, this route is not helpful. A better approach is to fit a variance profile model to the residuals $r_i$ themselves by minimizing the -2 log likelihood

$$L_r = \sum_{i=1}^{n} \left[ \frac{r_i^2}{\sigma_r^2 + \kappa_r^2 X_i^2} + \ln(\sigma_r^2 + \kappa_r^2 X_i^2) \right] \qquad (9).$$

This approach also permits a check for the special cases of constant variance ($\kappa=0$) and constant coefficient of variation ($\sigma=0$). This is done by evaluating $L_r$ with $\sigma$ set to the pooled standard deviation and $\kappa=0$; and with $\kappa$ set to the pooled coefficient of variation and $\sigma=0$. The difference between these $L_r$ values and that obtained from the minimum of (9) can be tested using a chi-squared with 1 degree of freedom. If this test concludes that $\kappa$ could plausibly be zero, then one could switch to the simpler unweighted Deming model, and if it concluded that $\sigma$ could plausibly be zero, then one could switch to the constant CV weighted Deming model. This fitting, and the tests, are performed in the PWD_resi function.

*Outliers*

As Deming regression is an L2 methodology, it can be derailed by severe outlier(s). A first step is some graphical checks of the approximately independent standard normal scaled residuals, but formal tests are an important backup for this visual inspection. This is particularly important if there are multiple outliers, as these can elude graphical checks because of "masking" (a situation in which outliers collude to hide each other) and "swamping" (where outliers collude to make good data points seem aberrant) [6].

While proper inferentials are elusive, if the sample is moderate to large, we can use an adaptation of Rosner's externally studentized deviation method. The approach implemented in the R function PWD_outlier is:

- Select a maximum number of outliers K to test for. The exact value if K is not critical but must exceed the actual number of outliers. A sensible default is 5% of *n*.

- Analyze the full sample and compute the scaled residuals.

- Select the case with the largest absolute scaled residual, and set it aside.

- Refit the model to the remaining *n*-1 cases. Find the case with the largest absolute scaled residual and add it to the set-aside list.

- Repeat the fitting and trimming operation until *K* cases are set aside. At this stage, you anticipate that the *n-K* retained cases are all "clean" and the outliers, if any, are among the *K* suspect readings that have been set aside.

Now explore reinclusion.

- Use the model fitted to the *n-K* clean cases to predict the *K* suspect cases. The difference between each actual *Y* and this prediction is a "predicted residual". Scale it using the precision profile model fitted to the residuals of the clean cases. Ignoring parameter uncertainty, these scaled predicted residuals should be standard normal. Find their two-sided P value under the standard normal distribution. Multiply this P value by *n-K* to turn it into a Bonferroni outlier P value. If all these Bonferroni P values are below the significance threshold, stop, and declare that all *K* suspect readings are significant outliers.

- If not, remove the case with the largest Bonferroni P value from the suspect list and add it to the "clean" set. Refit the model to the now *n-K+1* putatively clean cases and get the two-sided P value of the *K*-1 still under suspicion as before. Multiply this by *n-K+1* to calculate Bonferroni P values. Identify the case with the largest Bonferroni P. If this is significant, stop, and declare all cases still on the suspect list to be significant outliers. If not, move it from the suspect to the clean list.

- Continuing in this way, you will either find a significant Bonferroni P, or make your way right back to the first trimmed observation without getting one. If the latter, then declare that there are no significant outliers.

While technically imperfect, this algorithm should provide good protection against even troublesome scenarios with masking or swamping provided the sample is not too small.

*Sensitivity to model settings*

As noted earlier, the maximum likelihood estimates of $\alpha, \beta$ are unchanged if $g$ and $h$ are multiplied by an arbitrary constant – what matters is the shape of the ratio $g/h$ and not the actual functions. A further simulation exploring this was based on three 25 (OH) Vitamin D assays labeled 1, 2 and 3 whose precision profiles are well approximated by the RL model with $\sigma$ values of 0.5478, 0.8944 and 1.1392, and $\kappa$ values of 0.0247, 0.0831 and 0.0400 respectively. Samples of size 100 with true values in geometric progression from 8 to 80 ng/mL were used. This interval, an order of magnitude in size, contains two common medical decision levels, 12 and 30 ng/mL.

10,000 data sets were generated comparing each instrument with itself, and with each other instrument. Each such sample pair was analyzed by:

- A utopian fit, maximizing the likelihood over $\alpha, \beta$ using the true instrument precision profile,
- Passing Bablok regression,

- The Linnet constant coefficient of variation fit with $\lambda=1$ (labeled "L 1"),
- Where different instruments were being compared, the Linnet fit using as $\lambda$ the average of $g/h$ across the range of the samples (labeled "L gen")
- The RL fit assuming $\lambda=1$.

The intercept and slope of each fit, and the fitted value at the MDL of 12 ng/mL were computed, and their root mean squared error (RMSE) for estimating the true intercept 0, the true slope 1 and the true MDL 12. These RMSEs were then used to compute statistical efficiencies – the square of the ratio of the utopian RMSE to that estimator's. The results shown in the first six trios of rows of Table 1, which also shows the mean of the X-Y correlations across the 10,000 samples.

The table contains insights on all four of the regression approaches, but we focus on those for RL. Almost without exception, it is the best of the four estimators. In all the six settings and for all three estimands, RL's efficiency is close to 100% except for the comparison of instrument 2 with the much more precise instrument 1. Here its efficiency for the intercept is only 77% and that for the slope only 75%, and it is bested by the general Linnet fit.

The last three rows of Table 1 sketch the implication of the wrong functional form of the precision profile model. Instead of $SD^2 = \sigma^2 + (\kappa\mu)^2$, they generate data sets using the linear SD model $SD = \sigma + \kappa\mu$ which, like RL, goes to constant standard deviation $\sigma$ near zero and constant CV $\kappa$ for large values. Analysis however is by the RL. These three rows' $\sigma, \kappa$ values match those of instruments 1-3. The loss of statistical efficiency over where the true and fitted precision profile functional forms match is barely perceptible – the mismatch between the real precision profile and that used in the fitting costs almost nothing. Experience with heteroscedastic data in conventional least squares is that it is important to weight, but not essential to get the weighting function exactly right, so long as the weights at the high and low end of the range are about right, as is the case in this simulation.

*Illustrations*

We illustrate the use of the procedures with two data sets. The first analysis is a comparison of two assays for 25(OH)Vitamin D. Previous precision studies for the predicate and the test assay [7,8] analyzed by VFP [4] gave the fits

$$g(\mu) = 3.792\text{e-}16 + 0.06911\mu^{1.27}$$

$$h(\mu) = 6.106\text{e-}02 + 7.019\text{e-}05\mu^{2.18}.$$

It is not obvious from these functions, but the test instrument is much more precise than the older predicate – the *g/h* ratio ranges from 12 to 31. The weighted Deming regression using these true profiles and the prediction of the MDL of 12 ng/mL is:

```
Pearson correlation 0.9476 Spearman 0.9332
Fitted weighted Deming regression
Coefficient   Estimate se      CI
Intercept     -3.955   0.605   (-5.151,-2.758)
Slope         1.105    0.024   (1.058,1.152)
MDL   12.000 prediction   9.303 CI   8.458   10.148
```

The analysis is pictured in Figure 2. The top row is a scatter plot and a plot of residuals versus X. The bottom left shows the scaled residuals using the supplied *g* and *h* functions. The horizontal axis uses the index rather than X to spread the points evenly so that those with small X do not plot on top of each other obscuring the details. The bottom right is QQ plot following [9].

The index plot of residuals has quite constant width and no curvature and the QQr test for normality gives a P value of 0.6929. Both plots are satisfactory.

There is a problem however. The standard deviation of the scaled residuals is not 1 as it should be. A 95% confidence interval confirms that this difference is highly significant:

```
Scaled residual SD and CI   2.520   2.253   2.860
```

This situation is not uncommon – the assay variability measured in a precision study frequently fails to account for all the real-world variability of the MC. Carroll and Ruppert [10] discussed this problem in the unweighted Deming setting and pointed out that there are often additional sources of random variability that the historical precision study does not capture. They also showed that this under-accounting for variability biases the slope upward, a hint that the slope of 1.105 may be too high.

We can explore this further by fitting a RL precision profile to the scaled residuals: this should yield $\sigma_r$ =1, $\kappa_r$ =0; the actual estimates are $\sigma_r$ =2.30, $\kappa_r$ =0.0221. The formal tests for constant sd and CV give P values of 0.3297 and 0 respectively, indicating that the scaled residuals have a standard deviation higher by a constant factor of 2.3 than the published precision profiles imply.

However the lack of evidence of a non-zero $\kappa$ suggests that the prior information has the shapes of the precision profiles about right, even if the absolute magnitude was off target, and the regression coefficients can be trusted.

Next the data set was analyzed using the standalone analysis ignoring the precision profile information and using the default $\lambda$ =1 getting the variance profile estimates $\sigma$ = 0.8441, $\kappa$ = 0.0809 and coefficients:

```
Parameter estimate       se        CI
Intercept    -3.121    0.606 (-4.320,-1.923)
slope         1.086    0.024 ( 1.038, 1.134)
MDL   12.000 prediction    9.910 CI    9.052   10.768
```

The slopes and intercepts of the unknown and the known profile fits differ by about a standard error. It is perhaps surprising that they are that close when the assumption of a common RL variance profile is so far from true in both size and shape.

Suppose that while we did not have a good picture of either $g$ or $h$, we thought, based on general knowledge of these two assays, that $g/h$ was probably around 25. Rerunning using the RL model with $\lambda=25$ gives

```
Parameter estimate       se        CI
Intercept    -3.739    0.549 ( -4.824, -2.653)
slope         1.096    0.024 ( 1.050,  1.143)
MDL   12.000 prediction    9.416 CI    8.648   10.185
```

The slope is almost identical to that using the external $g$ and $h$ and the intercept differs by about a third of a standard error. The MDL is now quite a close match to that with known $g,h$. Despite the modeled shape being far from the truth, getting a rough idea of the assays' relative precision was enough to get a good answer.

As a second example, the first data set from the EP09 guidance [1] was analyzed. Its results are graphed in Figure 2. The regressions gave profile parameter estimates of $\sigma=0.1815$, $\kappa=0.0697$. The resulting fit is:

```
Parameter estimate       se        CI
Intercept    -0.129    0.319 (-0.763,0.505)
slope         1.030    0.023 ( 0.983,1.076)
```

This fit is very close to that given by Linnet's constant CV approach [11, 12] as one would expect given that the estimated $\sigma$ is close to zero and the fitted line is close to the identity.

The EP09 guidance suggested Passing Bablok regression as the appropriate method. This gave

```
Coefficient   Estimate       CI
Intercept     -0.089    (-1.129,0.940)
Slope          1.034    (0.958,1.104)
```

The precision profile weighted Deming estimates agree with those of Passing Bablok, but their confidence intervals are narrower. This is not unexpected since maximum likelihood has higher statistical efficiency than nonparametric methods which should translate into narrower confidence intervals.

The RL parameters for the residuals are $\sigma_r = 0.4316$ $\kappa_r = 0.1464$ and the resulting diagnostic plots are shown in Figure 3. They are unremarkable.

To illustrate outlier analysis, the first and last observations (1 and 100) were shifted upward and downward respectively. This bad placement leads to masking, as they collude to shift the slope down and the intercept up. It also widens both ends of the precision profile distorting the $\sigma, \kappa$ parameters.

The fitted RL model has $\sigma = 0.4427$, $\kappa = 0.0780$, both larger than the original data set. The regression shows the expected downward shift in the slope, by more than a standard error.

```
Parameter  estimate      se         CI
Intercept    0.340     0.573  (-0.797,1.476)
slope        1.003     0.035  ( 0.934,1.071)
```

The sequential outlier method with K=5 (namely 5% of the sample size) gave the initial suspects and their serial scaled residuals:

```
Suspect 100 outlier Z -3.29

Suspect  44 outlier Z  3.26

Suspect  74 outlier Z  3.12

Suspect   1 outlier Z  2.51

Suspect  40 outlier Z  2.27
```

Masking and swamping are apparent -- one of the two induced outliers was first to be selected, followed by two valid data and only then the other induced outlier. Backward reinclusion gave:

```
Least suspect   40 Z 2.261 BonP 2.3747

Least suspect   74 Z 3.185 BonP 0.1433

Least suspect   44 Z 3.607 BonP 0.0304
```

ending with the two induced outliers.

```
   case outlier Z Bonferroni P

1    1       4.200       0.00259

2  100      -3.897       0.00943
```

Refitting the model to the 98 retained cases gave the estimates

```
Parameter estimate      se         CI

Intercept   -0.144   0.417 (-0.971,0.683)

slope        1.032   0.027 ( 0.978,1.086)
```

These are essentially the same as those given by the original uncontaminated sample.

Figure 4 shows the results. The induced outliers are plotted as filled boxes. Case 1 is not striking in either of the plots in the top row. The lower left plot shows how case 1 is masked and case 44 swamped in the scaled residuals using the full sample. The right plot comes from the fit to the cleaned data. Here the two outliers are seen clearly in the index plot – note the wider range on the y axis.

One attraction of Passing Bablok regression is its outlier resistance. This is illustrated by the analysis of the outlier-contaminated sample:

```
Passing Bablok fit
Intercept   -0.0242   CI   -1.1510    1.1174
slope        1.0313   CI    0.9633    1.0924
```

These hardly differ from the original uncontaminated estimates.

*Conclusions*

Weighted Deming regression is potentially the most precise method for errors in variables regression, but achieving that potential requires appropriate weights. A constant CV model overweights low-concentration pairs and performs poorly if the measurement standard deviation fails to go to zero at low concentrations, and Linnet's constant CV algorithm, with its implicit assumption that the true regression is close to the line of identity, is biased for non-identity regressions. The R codes provided here implement weighted Deming with known precision profiles, and also cover the situation where the precision profiles' functional form, but not exact values, are known. The Rocke-Lorenzato functional form fits many actual data sets and is implemented in the codes. Other functional forms could be accommodated by recoding, some, like the linear standard deviation model, with only minor difficulty.

Simulation analysis of sensitivities based on actual precision profiles demonstrates that the RL profile model can be more precise than Linnet or Passing-Bablok, and there is evidence of robustness even when the actual precision profile differs from that being modeled algorithmically. As with conventional linear regression, it should be no surprise that precision profile models that are accurate for high and for low concentrations but less so for intermediate ones would work well, as these simulations illustrate.

The residuals and outlier assessment tools are particularly valuable byproducts of the parametric modeling.

| Intercept | | | | | | | | | |
|---|---|---|---|---|---|---|---|---|---|
| Test | mean r | Utopia | P_B | | L 1 | | L gen | | R_L | |
| | | RMSE | RMSE | eff | RMSE | eff | RMSE | eff | RMSE | eff |
| 1 1 | 0.9973 | 0.2050 | 0.2228 | 84.7 | 0.2393 | 73.4 | | | 0.2066 | 98.5 |
| 2 2 | 0.9758 | 0.4900 | 0.5468 | 80.3 | 0.5203 | 88.7 | | | 0.5016 | 95.4 |
| 3 3 | 0.9917 | 0.3919 | 0.4253 | 84.9 | 0.4862 | 65.0 | | | 0.3955 | 98.2 |
| 1 2 | 0.9865 | 0.3775 | 0.4602 | 67.3 | 0.4836 | 60.9 | 0.4035 | 87.5 | 0.4302 | 77.0 |
| 1 3 | 0.9945 | 0.3187 | 0.3559 | 80.2 | 0.4538 | 49.3 | 0.3838 | 69.0 | 0.3346 | 90.7 |
| 2 3 | 0.9837 | 0.4507 | 0.5001 | 81.2 | 0.5044 | 79.8 | 0.5598 | 64.8 | 0.4617 | 95.3 |
| 4 4 | 0.9952 | 0.2745 | 0.2981 | 84.8 | 0.3176 | 74.7 | | | 0.2777 | 97.7 |
| 5 5 | 0.9654 | 0.6462 | 0.7063 | 83.7 | 0.7312 | 78.1 | | | 0.6629 | 95.0 |
| 6 6 | 0.9850 | 0.5139 | 0.5543 | 85.9 | 0.6326 | 66.0 | | | 0.5207 | 97.4 |

| Slope | | | | | | | | | | |
|---|---|---|---|---|---|---|---|---|---|---|
| Test | mean r | Utopia | P_B | | L 1 | | L gen | | R_L | |
| | | RMSE | RMSE | eff | RMSE | eff | RMSE | Eff | RMSE | eff |
| 1 1 | 0.9973 | 0.0084 | 0.0090 | 86.9 | 0.0099 | 72.4 | | | 0.0085 | 98.7 |
| 2 2 | 0.9758 | 0.0243 | 0.0271 | 80.1 | 0.0252 | 92.5 | | | 0.0242 | 100.5 |
| 3 3 | 0.9917 | 0.0150 | 0.0159 | 89.1 | 0.0193 | 60.7 | | | 0.0151 | 98.9 |
| 1 2 | 0.9865 | 0.0183 | 0.0220 | 69.4 | 0.0213 | 73.9 | 0.0193 | 90.6 | 0.0192 | 91.2 |
| 1 3 | 0.9945 | 0.0123 | 0.0135 | 82.9 | 0.0180 | 46.8 | 0.0152 | 65.6 | 0.0128 | 92.8 |
| 2 3 | 0.9837 | 0.0203 | 0.0226 | 81.4 | 0.0224 | 82.4 | 0.0248 | 67.4 | 0.0204 | 99.2 |
| 4 4 | 0.9952 | 0.0111 | 0.0119 | 87.6 | 0.0131 | 71.7 | | | 0.0112 | 98.4 |
| 5 5 | 0.9654 | 0.0298 | 0.0320 | 86.5 | 0.0333 | 79.8 | | | 0.0296 | 101.1 |
| 6 6 | 0.9850 | 0.0199 | 0.0209 | 90.1 | 0.0255 | 60.7 | | | 0.0199 | 99.0 |

| MDL | | | | | | | | | | |
|---|---|---|---|---|---|---|---|---|---|---|
| Test | mean r | Utopia | P_B | | L 1 | | L gen | | R_L | |
| | | RMSE | RMSE | eff | RMSE | eff | RMSE | Eff | RMSE | eff |
| 1 1 | 0.9973 | 0.1347 | 0.1560 | 74.5 | 0.1478 | 83.0 | | | 0.1354 | 99.0 |
| 2 2 | 0.9758 | 0.2939 | 0.3493 | 70.8 | 0.3079 | 91.1 | | | 0.3074 | 91.4 |
| 3 3 | 0.9917 | 0.2616 | 0.3026 | 74.7 | 0.2981 | 77.0 | | | 0.2641 | 98.2 |
| 1 2 | 0.9865 | 0.2281 | 0.2860 | 63.6 | 0.2863 | 63.4 | 0.2389 | 91.1 | 0.2626 | 75.4 |
| 1 3 | 0.9945 | 0.2120 | 0.2490 | 72.5 | 0.2693 | 62.0 | 0.2375 | 79.7 | 0.2206 | 92.4 |
| 2 3 | 0.9837 | 0.2803 | 0.3297 | 72.3 | 0.3041 | 85.0 | 0.3248 | 74.5 | 0.2924 | 91.9 |
| 4 4 | 0.9952 | 0.1812 | 0.2106 | 74.0 | 0.1964 | 85.1 | | | 0.1833 | 97.8 |
| 5 5 | 0.9654 | 0.4080 | 0.4836 | 71.2 | 0.4394 | 86.2 | | | 0.4297 | 90.1 |
| 6 6 | 0.9850 | 0.3446 | 0.3991 | 74.6 | 0.3878 | 79.0 | | | 0.3511 | 96.4 |

Table 1

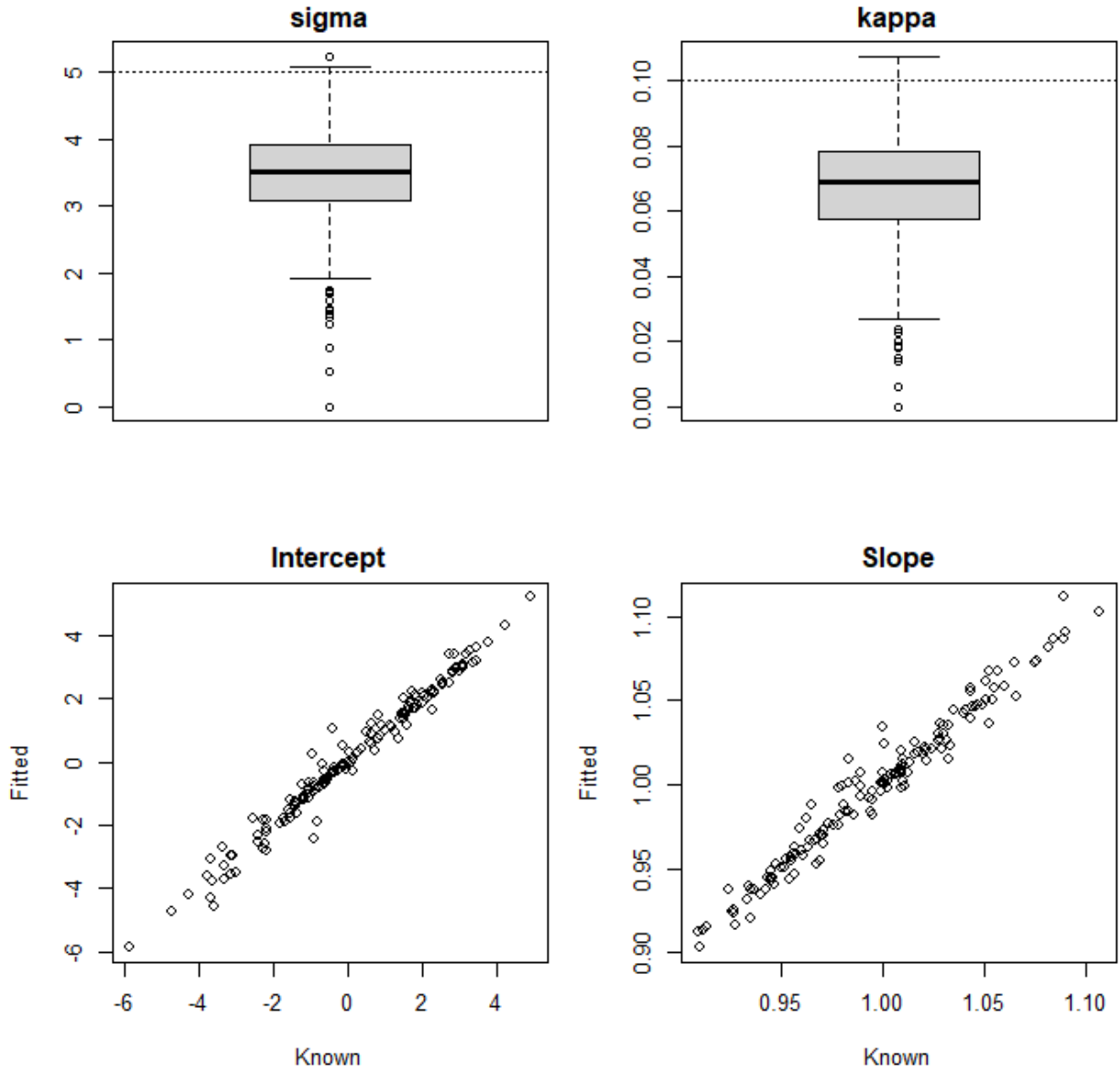

Figure 1. Estimates of four parameters.

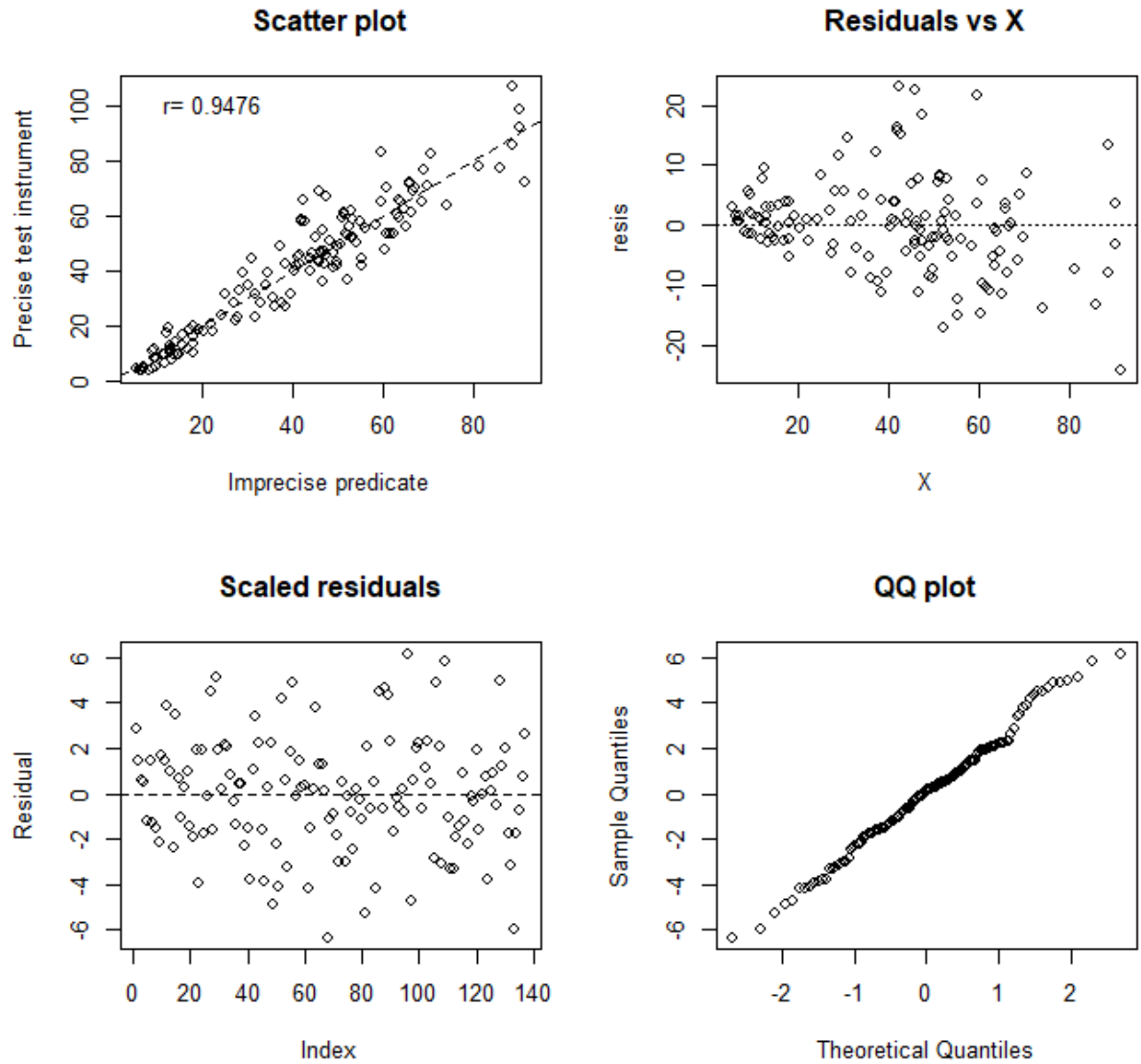

Figure 2. Analysis of Example 1 – Vitamin D

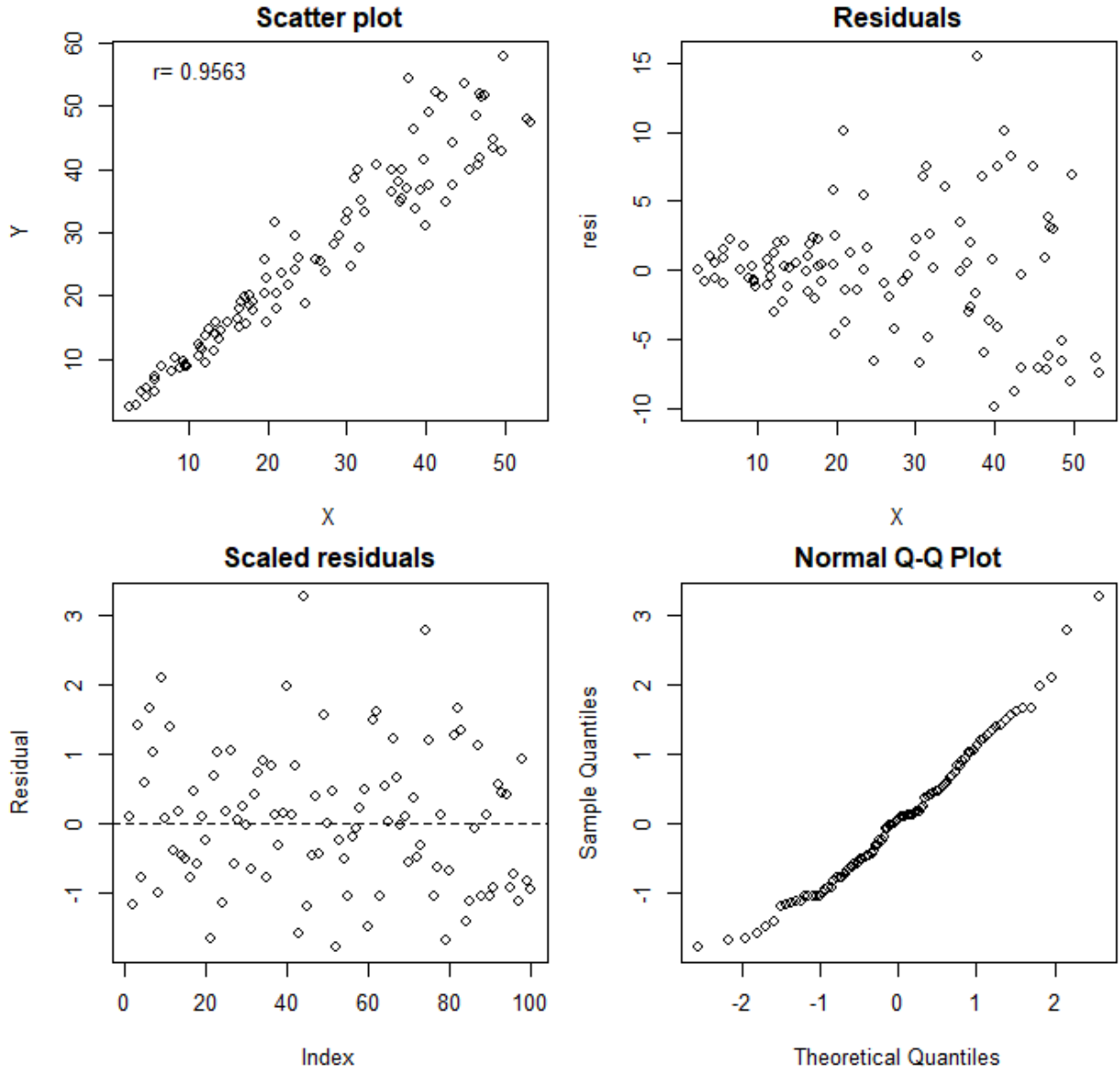

Figure 3.  Analysis of Example 2

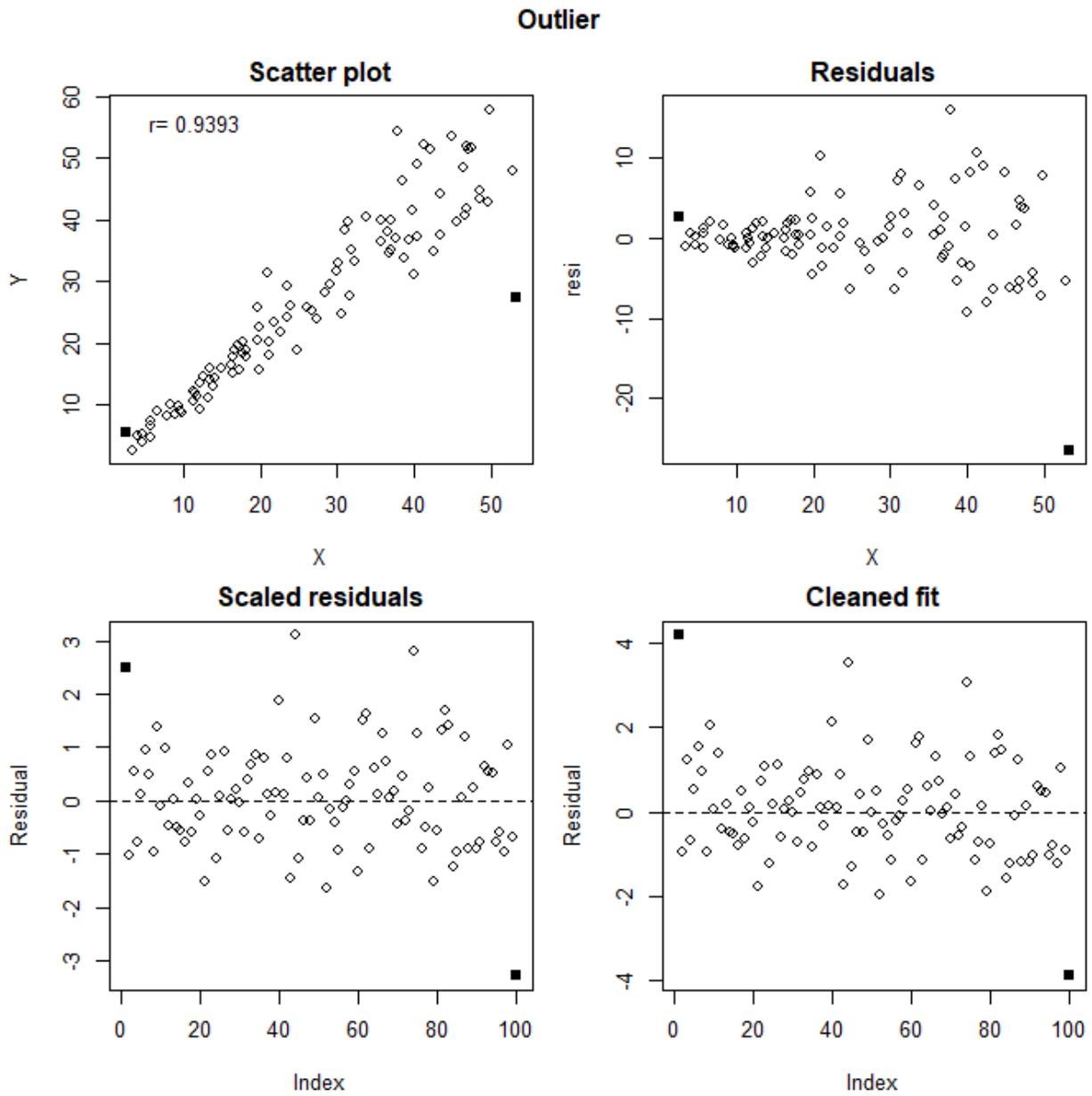
Figure 4. Analysis with two induced outliers

# Appendix
## A Critique of Linnet Constant CV Deming

If the errors of measurement are Gaussian, the assumption that both assays have constant CV leads to the distributions:

$$X_i \sim N(\mu_i, \lambda A \mu_i^2)$$
$$Y_i \sim N(\alpha + \beta \mu_i, A\{\alpha + \beta \mu_i\}^2) \qquad i = 1, 2, \ldots, n \qquad (A1)$$

This special case arises when the $\sigma$ of equation 8 is zero. Dropping it, removing the irrelevant $A$ and the now-irrelevant $\kappa$, and moving the scaling constant $\lambda$ from the $X$ term to the $Y$ term gives

$$L = \sum_{i=1}^{n} \left[ \frac{(X_i - \mu_i)^2}{\mu_i^2} + \frac{\lambda(Y_i - \alpha - \beta\mu_i)^2}{(\alpha + \beta\mu_i)^2} \right] \qquad (A2)$$

Maximum likelihood (ML) involves minimizing $L$.

Abbreviate the X and Y variances to $g_i = \mu_i^2$; $h_i = (\alpha + \beta\mu_i)^2 / \lambda$. For ML, the squared deviation in $X$ should be weighted by $1/g_i$ and that in $Y$ should be weighted by $1/h_i$.

Linnet forms a single consensus between the means of $X$ and $Y$, $c_i = (\mu_i + \alpha + \beta\mu_i)/2$ and weights the $X$ term by $1/c_i^2$ and the $Y$ term by $\lambda/c_i^2$, implicitly using the model

$$L = \sum_{i=1}^{n} \left[ \frac{(X_i - \mu_i)^2}{c_i^2} + \frac{\lambda(Y_i - \alpha - \beta\mu_i)^2}{c_i^2} \right]; \quad c_i = \{\alpha + (\beta+1)\mu_i\}/2 \qquad (A3)$$

This is the same as (2) if $(\alpha, \beta) = (0,1)$, but otherwise the weights implied by (A2) and by (A3) can be arbitrarily different.

Turning to the $\mu_i$, if $\alpha$ and $\beta$ are given, then the MLEs of the latent true concentrations and the true means of $X_i$ and $Y_i$ under the model (A1) are

$$\hat{\mu}_i = \frac{h_i X_i + \lambda\beta g_i (Y_i - \alpha)}{h_i + \lambda\beta^2 g_i}; \quad \alpha + \beta\hat{\mu}_i \qquad (A4)$$

in which, as $\mu$ is on both sides of the equation, iteration is needed.

Linnet uses angular projection to the line to define
  $X_i - \lambda\beta(Y_i - \alpha - \beta X_i)/(1 + \lambda\beta^2)$ as the estimated mean for $X_i$ and
  $Y_i - (Y_i - \alpha - \beta X_i)/(1 + \lambda\beta^2)$ as the estimated mean for $Y_i$

If the correlation is high and the true line is close to the identity, there is little ambiguity in the $\mu_i$ so both approaches should be similar, but it is not clear how well this holds otherwise.

To test this more general situation, I generated 10,000 samples of size 120 by model (1) with true concentrations going in geometric progression from 1 to 10. The true regression line had intercept 2 and slope 1.5 and both *X* and *Y* had 20% coefficient of variation. Figure A1 shows one of the simulated data sets with the true generating line, along with the ML and Linnet fits.

Figure A2 is a pair of comparative box and whisker plots of the intercept and slope produced by the Linnet and ML algorithms. The plots show that the Linnet method is severely biased for both the intercept and slope, which is not the case with the ML.

Across the 10,000 simulations, the root mean squared error of the estimates was:

Root mean squared error

| Intercept | | Slope | |
|---|---|---|---|
| Linnet | ML | Linnet | ML |
| 0.8082 | 0.2533 | 0.2355 | 0.0932 |

This table quantifies the much worse performance of the Linnet algorithm.

A coding check repeated the simulation using the line of identity as the true line. Here the two algorithms gave essentially the same results with Linnet having a slight edge – see Figure 3.

Root mean squared error

| Intercept | | Slope | |
|---|---|---|---|
| Linnet | ML | Linnet | ML |
| 0.1120 | 0.1259 | 0.0500 | 0.0535 |

The conclusion is that the Linnet algorithm probably works well in method comparisons where the true regression line is close to the line of identity and the correlation is high, but may be deficient otherwise. Arguably, it should be replaced by a ML implementation following (2), which should work regardless of the intercept and slope of the true line.

*Reference*
Linnet K. Estimation of the linear relationship between the measurements of two methods with proportional errors. Stat Med. 1990; 9: 1463-1473

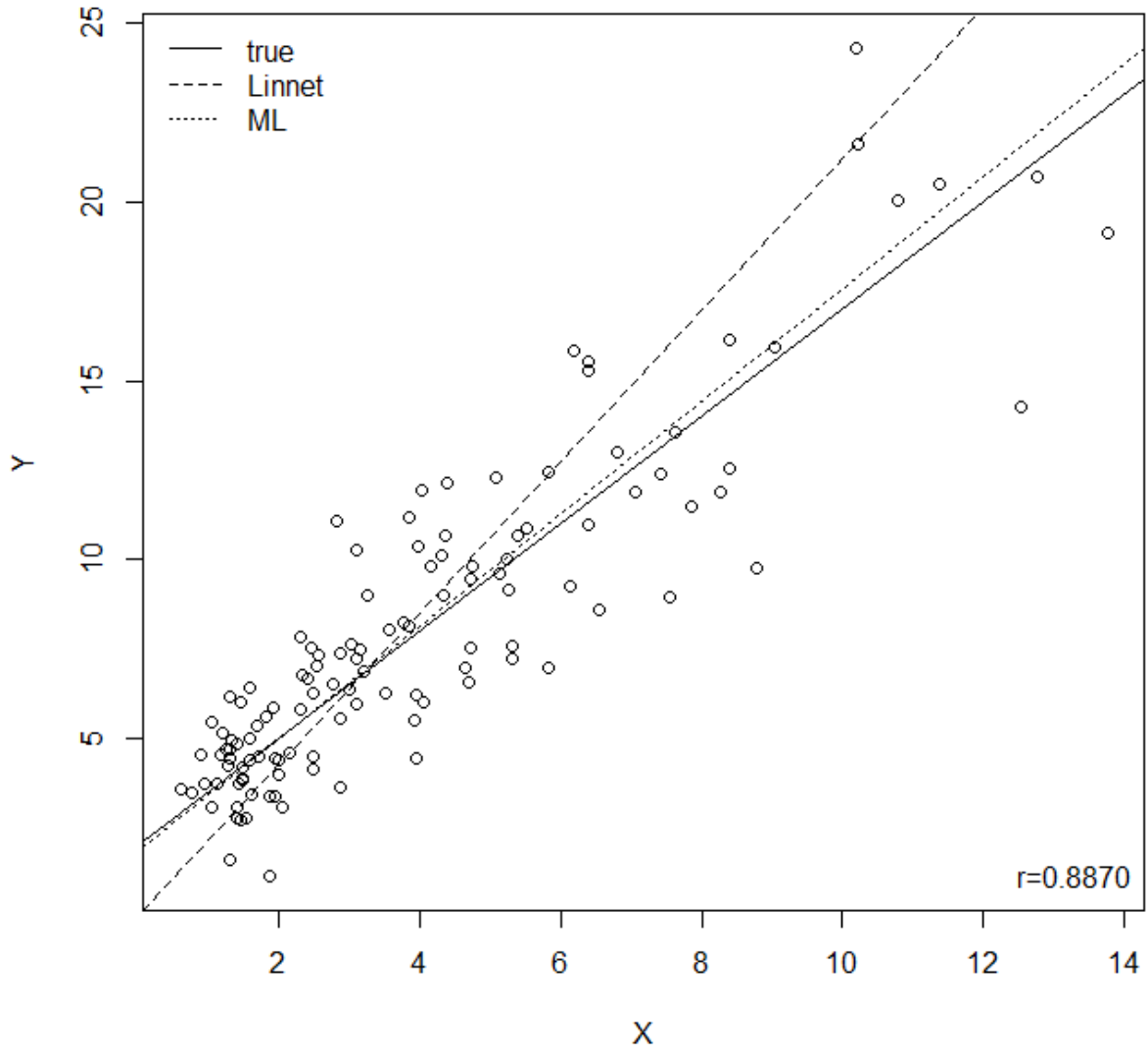

Figure A1. A simulated data set

*Precision profile weighted Deming methods comparison* 29

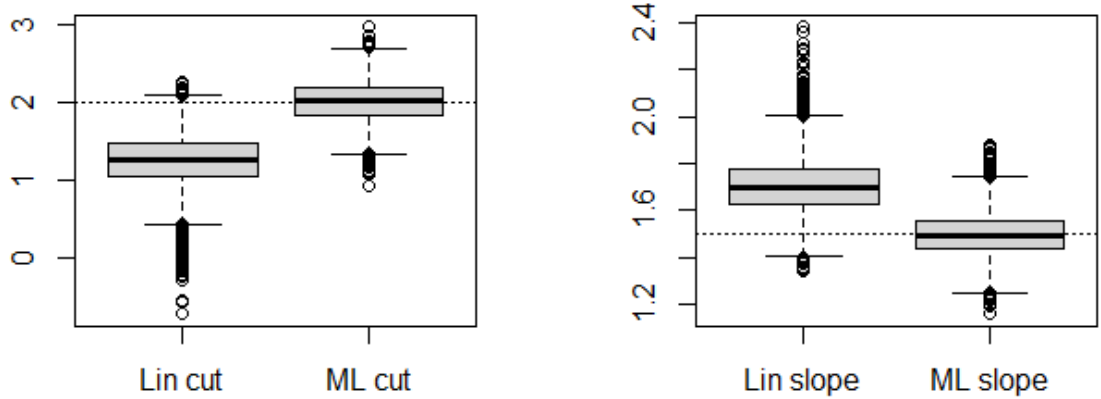

Figure A2. Linnet and ML non-identity true regression.

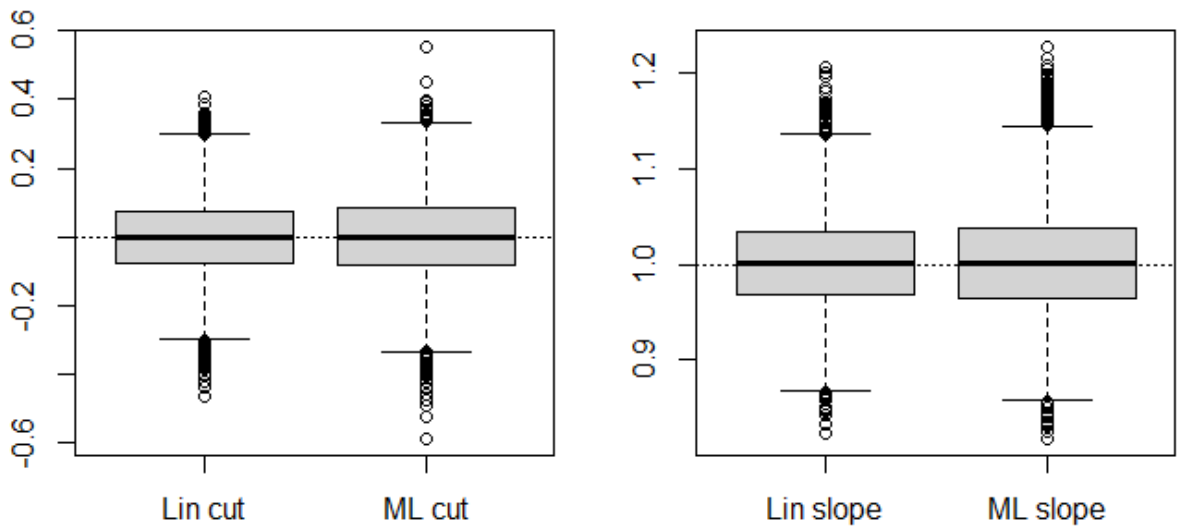

Figure A3. Counterpart for the identity.